\documentclass[final,preprint,amsmath,amssymb,amsfonts,showpacs,superscriptaddress,titlepage,nobalancelastpage,raggedbottom,floatfix]{revtex4}

\usepackage[english]{babel}
\usepackage{ifpdf}
\ifpdf
\usepackage[pdftex]{graphicx}
\usepackage[protrusion=true,expansion=true]{microtype}
\usepackage[pdftex,colorlinks,linkcolor=black,citecolor=black,urlcolor=black,hyperfootnotes=false]{hyperref}
\else
\usepackage[dvips]{graphicx,color}
\usepackage{microtype}
\fi

\newcommand{\remove}[1]{}

\begin{document}
\selectlanguage{english}
\title{Designing colloidal ground state patterns using short-range isotropic interactions}
\author{Simon H. Tindemans}
\email{tindemans@amolf.nl}
\author{Bela M. Mulder}
\email{mulder@amolf.nl} \affiliation{FOM Institute AMOLF, Science
Park 104, 1098 XG, Amsterdam, The Netherlands}
\date{\today}
\begin{abstract}
DNA-coated colloids are a popular model system for self-assembly through tunable interactions. The DNA-encoded linkages between particles theoretically allow for very high specificity, but generally no directionality or long-range interactions. We introduce a two-dimensional lattice model for particles of many different types with short-range isotropic interactions that are pairwise specific. For this class of models, we address the fundamental question whether it is possible to reliably design the interactions so that the ground state is unique and corresponds to a given crystal structure. First, we determine lower limits for the interaction range between particles, depending on the complexity of the desired pattern and the underlying lattice. Then, we introduce a `recipe' for determining the pairwise interactions that exactly satisfies this minimum criterion, and we show that it is sufficient to uniquely determine the ground state for a large class of crystal structures. Finally, we verify these results using Monte Carlo simulations.
\end{abstract}

\pacs{82.70.Dd, 81.16.Dn, 61.50.Ah, 87.14.gk}

\maketitle

%%%%%%%%%%%%%%%%%%%%%%%%%%%%%%%%%%%%%%%%%%%%%%%%%%%%%%%%%%%%%

\section{Introduction}

Micrometer-sized colloidal particles have the desirable property
that they are small enough to undergo thermal motion, yet large
enough to be accessible to direct manipulation and controlled
fabrication. This makes them into an ideal class of systems to
investigate the relation between particle interactions and the
large-scale properties of the structures they form. Rather than
the traditional approach of determining the nature of the
interactions between particles in existing materials and
reconstructing the material properties from these interactions, we
can now attempt to use tailor-made colloidal particles to
\emph{design} new materials. The question then becomes whether the
interactions between particles can be chosen in such a way that
they form a system with the required emergent properties. Because
the desired interactions (inputs) are determined for a given
system (output), this approach has also been named `inverse'
statistical-mechanics (see \cite{Torquato2009} for a recent
overview).

\begin{figure}[ht]
\begin{center}
\includegraphics{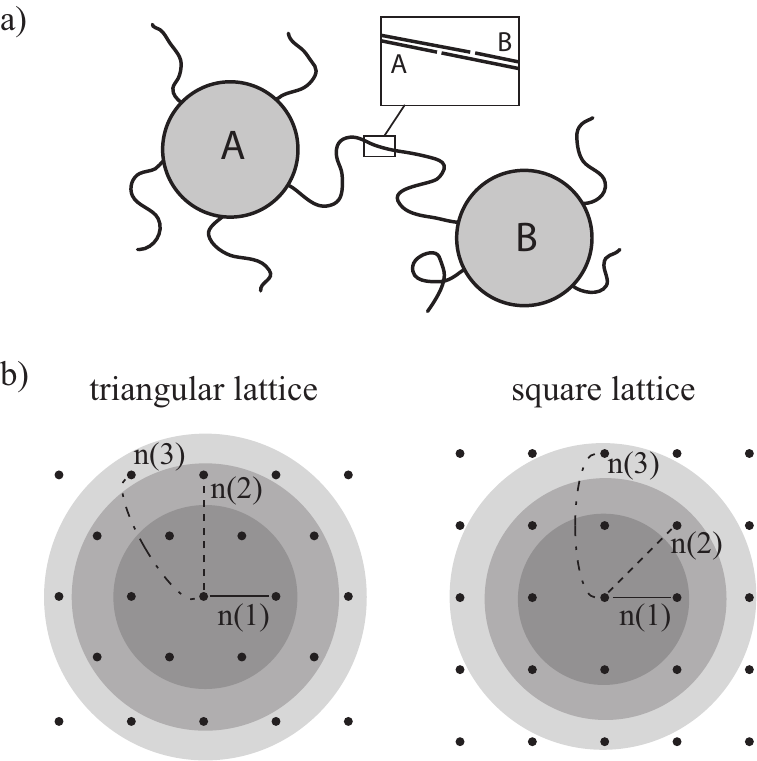}
\caption{a) Schematic overview of colloid-colloid binding through
DNA interactions. b) Overview of the nearest neighbor ($n(1)$),
next-nearest neighbor ($n(2)$) and next-next-nearest neighbor
($n(3)$) interactions on triangular and square lattices.}
\label{fig:colloid-lattice}
\end{center}
\end{figure}

A promising model system for particles with extensively tunable
interactions is that of DNA-coated colloids. In this system, first
proposed by Mirkin \emph{et al.} \cite{Mirkin1996}, DNA is grafted
onto micro-meter-sized colloids. The two strands forming the DNA
are designed to differ slightly in length, so that a short stretch of
single stranded DNA is exposed at the far end of the strand. This
creates what is called a `sticky' end, to which another single DNA
strand can bind through hybridization. The multitude of possible
binding sequences provided by DNA's four-letter alphabet combined
with the specificity of base pair binding means that the binding
affinity can be precisely manipulated. This specificity is not limited to a
single pair of DNA strands, but pairwise-specific interactions can be
defined simultaneously for a large number of sticky ends. In
principle, it is possible to create a system with many different
types of colloids, each of which has tunable interactions with
every other type of colloid. The sticky ends of two different
colloids can either bind directly (see figure
\ref{fig:colloid-lattice}), or mediated by a piece of linker DNA
that has preferential affinities for both types of colloids. The
latter provides more flexibility to control the interactions in
the system, especially when many colloids with different sticky
ends are involved.  The ultimate goal of the research on
DNA-coated colloids is to design materials that will, in the
proper circumstances, self-assemble from their constituents.

The range of possibilities offered by DNA-mediated interactions is
fundamentally different from electrostatic interactions, for which
the control is limited to the sign, strength and range of the
interactions without further possibilities for discrimination.
Patterns composed of charged colloids are therefore generally
limited to two colloidal species (see, for example
\cite{Leunissen2005}). Arguably, the potential of DNA-mediated
interactions has so far not even been fully utilized, as
experiments have only been done on systems with single
\cite{Biancaniello2005,Hill2008} or two species of colloids
\cite{Nykypanchuk2008,Park2008,Geerts08}. Also, theoretical work
\cite{Tkachenko2002} and simulations
\cite{Lukatsky2004,Bozorgui2008} have predominantly focused on
systems with two types of colloids. Notable exceptions are the
work by Licata and Tkachenko \cite{Licata2006}, which was limited
to a single particle per colloid type, and the lattice model by
Lukatsky \emph{et al.} \cite{Lukatsky2006} that used four species
of colloids, and provided the basis for this work.

In order for a collection of DNA-coated colloids to self-assemble
into a complex pattern, every colloid -- through its interactions
-- should contain enough information to `find' its target
position. A complication is that the distribution of the sticky
DNA ends that mediate the interactions is to a first approximation
isotropic \cite{Crocker2008} so that the interactions can only be
a function of the distance between colloids. Furthermore, for
practical purposes, this distance is constrained by the length of
the DNA \cite{Geerts08}. Therefore we may ask ourselves whether it
is at all possible to design large-scale patterns using only
short-range isotropic interactions, and, if so, whether there is a
lower limit to this interaction range. We address this question in
the context of a two-dimensional lattice model with isotropic
interactions and we investigate whether the interactions between
the colloids on this lattice can be designed in such a way that
the colloids self-assemble into a complex crystal structure.

A necessary requirement for self-assembly is that the target
structure represents the unique minimum energy ground state of the
system, so that it will be the preferred state for $T \to 0$. For
the basic theoretical work presented here we therefore restrict
ourselves to the design of a unique ground state for non-trivial
crystal structures, using only isotropic interactions.

After introducing the model, we derive a requirement for the
minimum interaction range between colloids, depending on the
symmetry of the underlying lattice and the size of the desired
pattern. Subsequently, we present a minimal recipe for the
interactions that allows for the design of arbitrarily large
periodic patterns. Besides periodic patterns, this recipe can also
be used to construct patterns with glide reflections and two-fold
rotations. Finally, these results are illustrated by means of
Monte Carlo simulations that demonstrate the self-assembly of the
designed patterns from random initial conditions.

\section{Model definition}

We define a simple geometrical lattice model for the interactions
in a system of DNA-coated colloids, as a generalization of the
model by Lukatsky \emph{et al.} \cite{Lukatsky2006}. Let us
consider a lattice on which every site may be occupied by at most
one colloid, corresponding to an excluded volume effect with the size of a lattice unit. Each colloid is of a particular type that is characterized
by its sticky end and labelled by an alphabetical index
(\texttt{A}, \texttt{B}, etc.). Formally, every lattice site $i$
is in a state
\begin{equation}
s(i) \in \{\mathtt{A, B, ... , Z} \} \cup \emptyset,
\end{equation}
in which $Z$ is the final element of the set of distinguishable
colloids (not necessarily containing 26 elements) and $\emptyset$
denotes an empty site. Empty sites can be regarded as an
additional non-interacting colloid species. We can trivially
assign a vector in a $Z$ dimensional vector space to each of the
possible states through the identification
\begin{equation}
\mathtt{A} = \left( \begin{array}{c}  1 \\ 0 \\  \vdots \\ 0
\end{array} \right), \quad \ldots, \quad \mathtt{Z} = \left( \begin{array}{c} 0 \\ \vdots \\ 0 \\1
\end{array} \right); \quad \emptyset = \left( \begin{array}{c} 0 \\ \vdots \\ 0
\end{array} \right) \label{eq:vectordef}
\end{equation}

The colloids on the lattice have DNA-mediated interactions that give rise to an
effective two-body interaction energy that is both isotropic and
short-ranged. This interaction will therefore be a function of the
distance on the lattice. It is convenient to introduce the shorthand notation $n(r)$, $r
\in \mathbf{N}$, for the set of particles with identical
site-to-site distances on the lattice, ordered by increasing
distance for increasing $r$. $n(1)$ thus contains all
nearest-neighbor pairs, $n(2)$ all next-nearest-neighbor pairs,
etc. In the context of this work, we restrict ourselves to
two-dimensional lattices, but the language used to describe the
interactions is also applicable to lattices in higher
dimensions. See figure \ref{fig:colloid-lattice} for the
interaction ranges for the square and triangular lattices used in
this work.

The Hamiltonian for the system is defined as
\begin{equation}
H = \frac{1}{2} \sum_{r = 1}^{R} \sum_{(i,j) \in n(r)}
J_{mn}^{(r)} s_m(i) s_n(j),\label{eq:hamiltonian}
\end{equation}
with an implicit summation over the vector indices $m$ and $n$.
$R$ indicates the maximum range of the particle interactions,
as measured along the lattice links, and we introduce a symmetric
interaction matrix $J^{(r)}$ for every lattice distance $r$. The
aim is to design the interaction matrices $J^{(r)}$ in such a way
that the Hamiltonian is uniquely minimized for a predetermined
crystal lattice -- up to global transformations corresponding to
the symmetries of the underlying lattice (rotations, translations
and reflections).

At this point, it will be clear that the above model can easily be
used to describe \emph{any} multi-state lattice model with
pairwise interactions that are both short-ranged and isotropic.
However, for conceptual clarity, we refer only to the
DNA-coated colloid system and leave further applications up to
imagination of the reader.

\section{Lower limits to the interaction range}\label{sec:rangelimits}

So far, the maximum interaction range $R$ has not been
specified. For practical applications this interaction range
should be as short as possible. However, it is reasonable to
expect that a very short interaction range may impose limitations
on size or complexity of the patterns that can be designed.
In this section we determine lower limits for the interaction
range, depending on the underlying lattice type (square,
triangular) and the desired complexity of the designed pattern.

The square and triangular lattices have a large number of intrinsic
symmetries: translations, rotations, reflections and glide
reflections (reflections accompanied by a translation along the
reflection axis). The collection of these symmetries is summarized
by the corresponding wallpaper groups: \emph{p4m} for the square
lattice, \emph{p6m} for the triangular lattice
\cite{Gruenbaum1987}. Applying any of these symmetry operations to
a configuration of colloids \emph{on} the lattice produces another
valid (on-lattice) configuration. Furthermore, for any given
configuration and symmetry operation, a number of
\emph{symmetrized configurations} can be constructed that are
invariant under the given symmetry operation. For example, a pair
of reflection-symmetrized configurations is created by copying the
pattern on one side of a reflection axis over to the other side, and
vice-versa.

The goal of this work is to design the interaction matrices in
such a way that the system has a unique ground state for a
configuration of colloids that is equal to an \emph{a priori}
specified pattern. Naturally, this means the energy of this
distribution should be lower than that of every other distinguishable
distribution, including all possible symmetrized forms. We
conclude that the ground state can only possibly be unique if each
of the possible symmetrized distributions either (a) has a larger energy or (b) is indistinguishable from the original distribution (i.e. it was
already symmetric).

\begin{figure*}[htp]
\begin{center}
\includegraphics{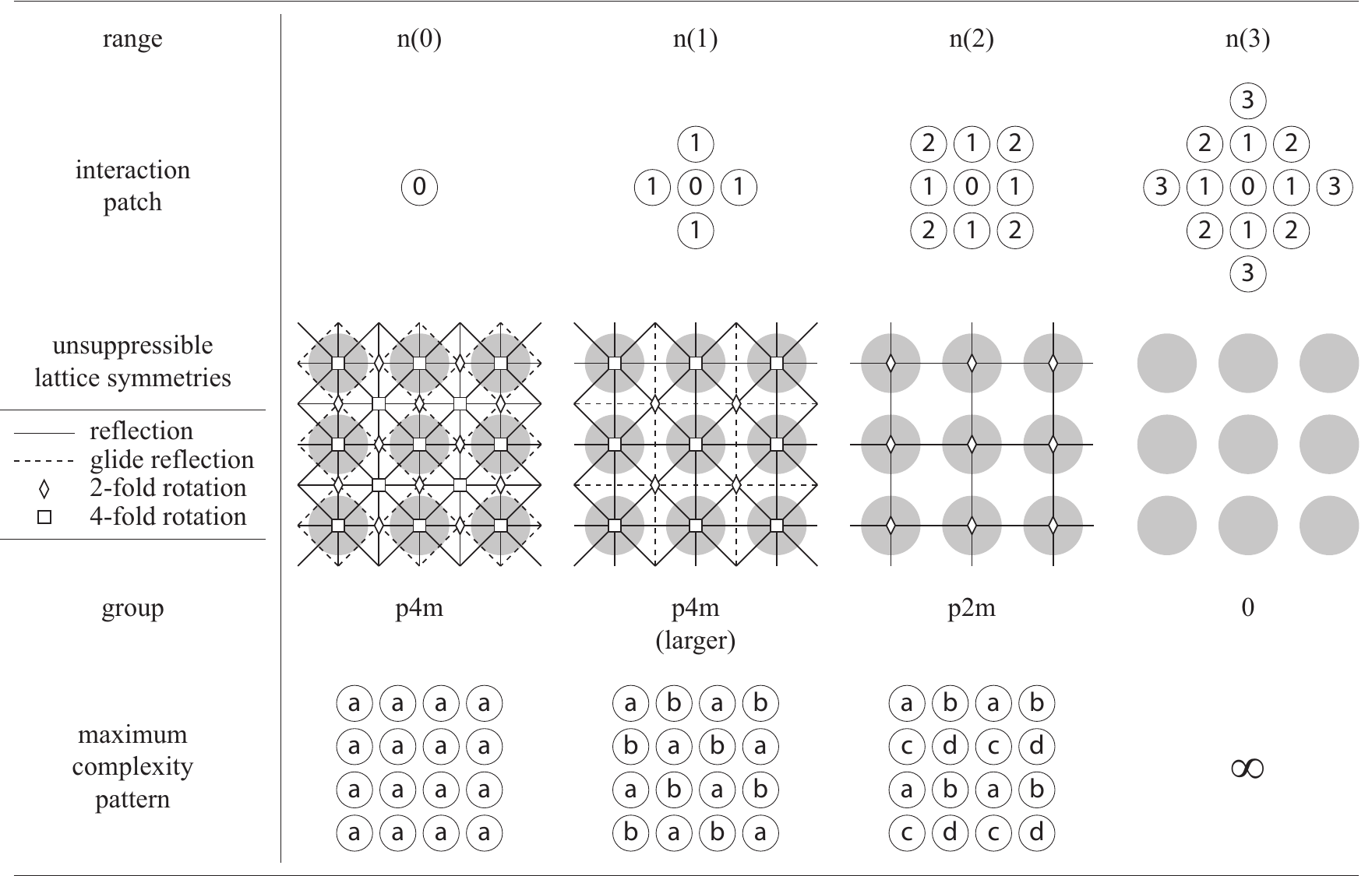}
\caption{Table relating the interaction range, the unsuppressible
lattice symmetries and the corresponding maximum complexity
patterns on a square lattice.} \label{fig:limit-square}
\end{center}
\end{figure*}

This observation gives us a handle on the relation between ground
state design and interaction range. Figure \ref{fig:limit-square}
gives a graphical overview of this relation on a square lattice.
Starting on the left, for an interaction range of $n(0)$ (no
interactions) the interaction energy is always zero. Obviously,
any symmetrized solution will have the same energy, so a solution
can only be a unique ground state if it itself is symmetric under
all symmetry operations on the lattice (\emph{p4m}). This is shown
as the maximum complexity pattern on the bottom left, consisting
of only \texttt{A}-type colloids.

Increasing the interaction range to $n(1)$, the nearest-neighbor
interactions along lattice links can be used to distinguish a
colloid from its neighbors, as shown in the interaction patch by
the different labels 0 and 1. The presence of these interactions
can cause a symmetry-generated state to have a higher energy than the
original state. Graphically speaking, this happens when, for a
given symmetry operation, the interaction patch can be placed on a
position such that either (a) the 0 is mapped onto a 1 or
vice-versa, or (b) the patch is split by a glide reflection. A
large fraction of the intrinsic lattice symmetries can be
suppressed this way, such as the reflection symmetry between
columns and rows of colloids. The remaining symmetries are shown
in the second row of figure \ref{fig:limit-square} (the glide
symmetries are a resultant of the remaining symmetries). Again,
the symmetry group is \emph{p4m}, but this time with a unit cell
of two lattice sites. Every ground state should by symmetric under
this group, so that the most complex ground state pattern is given
by the checkerboard pattern of \texttt{A} and \texttt{B} type
colloids.

Increasing the range yet again to $n(2)$ further reduces the
symmetries that cannot be suppressed to the \emph{p2m} wallpaper
group, corresponding to a $2\times 2 $ repeated pattern of
colloids. This class of interactions and the corresponding
solutions have been used by Lukatsky \emph{et al.}
\cite{Lukatsky2006}. When the interactions include $n(3)$, all
lattice symmetries can be suppressed and this method does not
indicate any remaining symmetry-derived limitations. An
interaction range of $n(3)$ or longer is therefore a necessary
condition for the design of patterns that are larger than $2
\times 2$. In the next section, we will show that this interaction
range is also sufficient.

\begin{figure*}[htp]
\begin{center}
\includegraphics{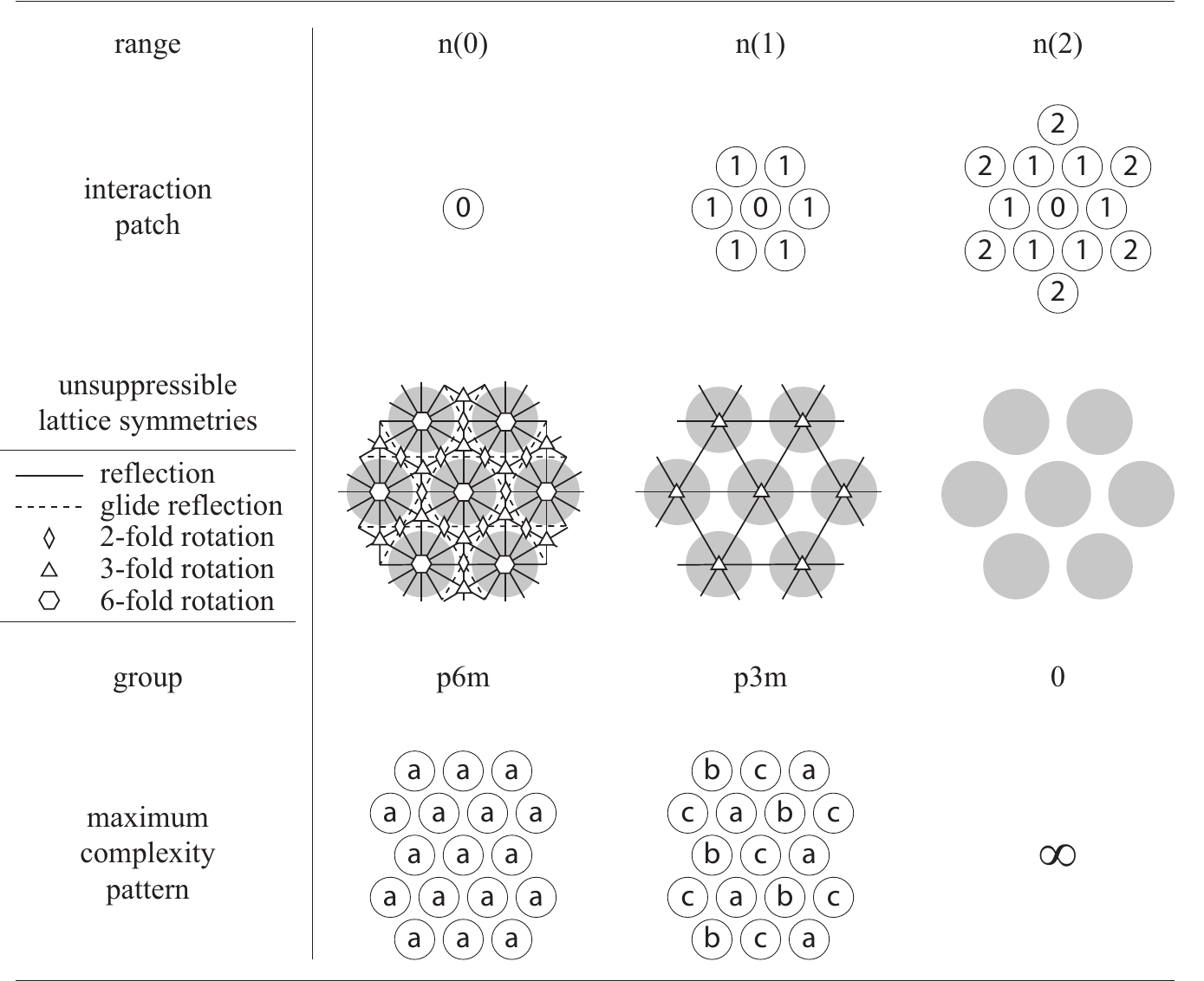}
\caption{Table relating the interaction range, the unsuppressible
lattice symmetries and the corresponding maximum complexity
patterns on a triangular lattice.} \label{fig:limit-hex}
\end{center}
\end{figure*}

Figure \ref{fig:limit-hex} shows a similar derivation on the
triangular lattice. Starting from the full symmetry of the lattice
(\emph{p6m}), symmetries are suppressed by increasing the
interaction range. For $n(1)$ interactions the remaining lattice
symmetries are of group \emph{p3m}, corresponding to a triangular
3-colloid pattern. In contrast to the square lattice, $n(2)$
interactions are already sufficient to suppress all lattice
symmetries on the triangular lattice. The design recipe that is
presented in the following section will prove that the $n(2)$
interaction range is not only necessary, but also sufficient for
designing unique ground states with arbitrarily large numbers of
colloids.

\section{Proof of designability}\label{sec:groundstate}

The results from the previous section indicate that non-trivial
patterns can only be reliably designed if interaction range is at
least $n(2)$ (on a triangular lattice) or $n(3)$ (on a square
lattice). Building on this generic result we now introduce a
simple recipe for interactions between the colloids that is both
minimal in terms of the interaction range and still guaranteed to
produce a unique ground state for a large class of patterns.

The prescription for the interactions consists of a mixture of
positive and negative design elements. First, each pair of
colloids that should form a nearest-neighbor pair in the final
pattern is assigned a negative interaction energy $-\alpha < 0$ in
the nearest-neighbor matrix $J^{(1)}$ (positive design). In
addition, we introduce a repulsive interaction at longer distances, represented by a
contribution $\varepsilon > 0$ for all diagonal elements in the
next-nearest neighbor matrix $J^{(2)}$ and, for square lattices, also
the next-next-nearest neighbor matrix $J^{(3)}$. The remaining
interactions are set to zero. This choice of the interaction
matrices is far from unique, but it is simple, leading to sparse
matrices, and it is sufficient to guarantee a unique ground state
with an energy of $-\alpha n/2$ per particle, where $n$ is the
coordination number (the number of nearest-neighbors per site) of
the lattice. A motivation and proof for this design strategy is
given below, for both lattice types under consideration.

\subsubsection{One dimension}

\begin{figure}[ht]
\begin{center}
\includegraphics{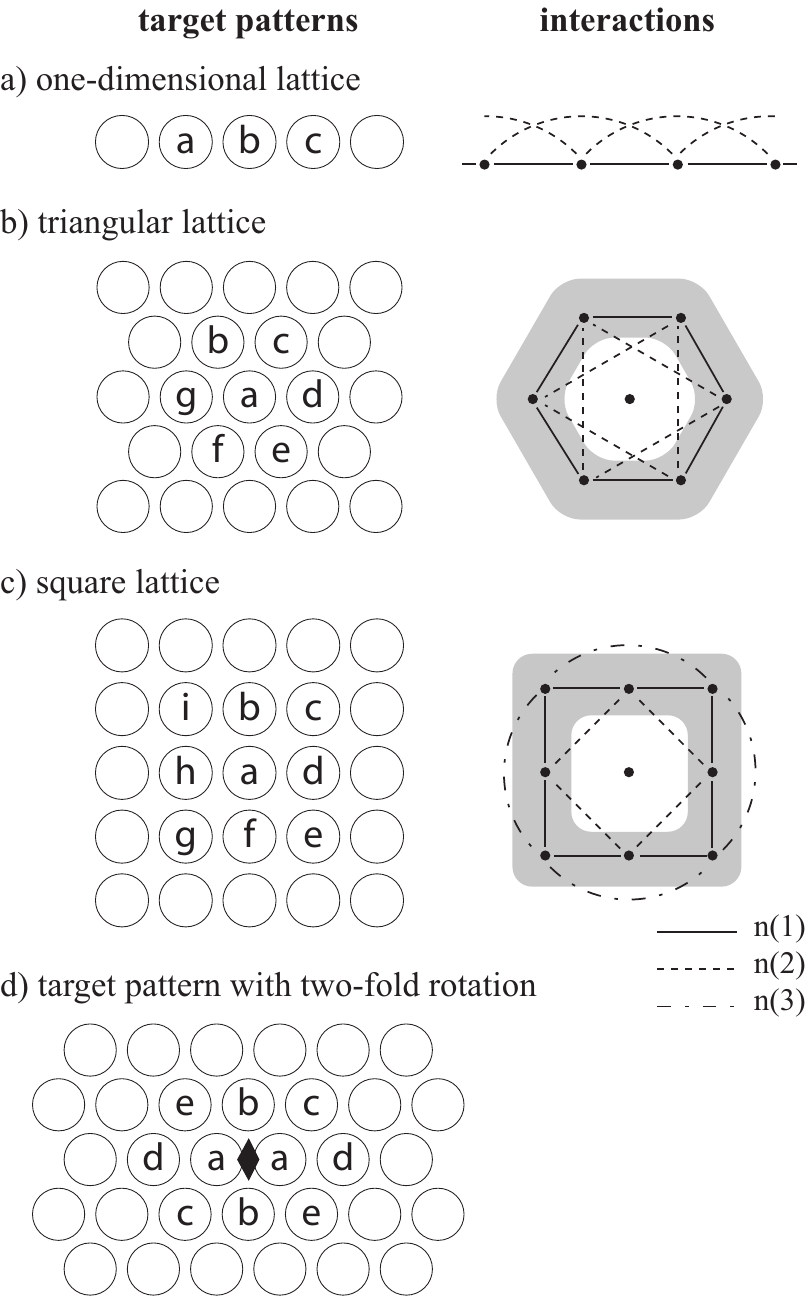}
\caption{Overview of the local target patterns for one- and two-dimensional lattices and corresponding interactions between lattice sites. The shaded areas in (b) and (c) indicate the mapping of local neighborhoods onto a one-dimensional periodic structure. Panel (d) depicts the local neighborhood of a two-fold rotation center on a triangular lattice.} \label{fig:patterns-interactions}
\end{center}
\end{figure}

It is instructive to initiate our analysis with a one-dimensional
system (see figure \ref{fig:patterns-interactions}a). The symmetry argument from the previous section suggests
that the interactions should range up to $n(2)$ in order to
suppress the reflection symmetry through the center of each
colloid. Suppose we wish to design a linear pattern containing the
sequence \texttt{...ABC...}. Using only nearest-neighbor
interactions this requires a preferential binding of \texttt{B} to
both \texttt{A} and \texttt{C}. However, when \texttt{B} is bound
to \texttt{A} on one side, there is nothing to stop another
\texttt{A} from binding on the other side of \texttt{L}, producing
the sequence \texttt{ABA}. Hence, as expected, the only sequence
that can reliably be designed using next-neighbor interactions
alone is an alternating sequence.

This restriction can be circumvented by extending the range of
interactions to $n(2)$. This allows us to include a
\emph{self-repulsion} at range $n(2)$ for both \texttt{A} and
\texttt{C}. Suppose \texttt{A} binds to \texttt{B} first, forming
the complex \texttt{AB}. The long-range self-repulsion of
\texttt{A} will prevent it from also binding to the other side of
\texttt{B}, or it can only do so with a reduced affinity. In both
cases, if \texttt{C} has approximately the same nearest-neighbor
affinity for \texttt{B} as \texttt{A} does, the ground state will
contain the sequence \texttt{...ABC...}, or its reverse, which is
allowed by a global symmetry operation. The same argument can be
applied iteratively to each position in the sequence, from which
we conclude that a ground state sequence of infinite length can
indeed be designed. Using the interaction recipe defined above, all repulsive interaction ($n(2)$) are avoided and only the attractive interactions ($n(1)$) remain. This way, every link in the lattice obtains an energy $-\alpha$, corresponding to the lowest possible energy state of the lattice, which
proves that it it indeed the ground state.

This argument shows that it is theoretically possible to design
arbitrarily long unique strings of letters, but it may be more
desirable to form a string of a limited length, say
\texttt{AB...YZ}, and to create a repeated `tiling' based on this
fragment. Implementing this periodic boundary condition is
straightforward, as we can simply instruct \texttt{A} and
\texttt{Z} to become nearest neighbors by setting their
interaction energy to $-\alpha$ (positive design).

\subsubsection{Triangular lattice in two dimensions}

We now address the more interesting case of two-dimensional
designs, starting with the triangular lattice. As in the
one-dimensional case, a minimum interaction range extending to the
next-nearest neighbors ($n(2)$) is required to design crystals
with more than three colloids (see section \ref{sec:rangelimits}).
That the design recipe is also sufficient can be seen as follows.

Consider a single particle \texttt{A}, having six
nearest-neighbors, and suppose we wish to design a ring of
particles \texttt{BCDEFG} around \texttt{A} (see figure
\ref{fig:patterns-interactions}b, left). Attractive nearest-neighbor interactions
between colloid \texttt{A} and the other particles cause the ring
of particles around \texttt{A} to be made up of the correct
colloids in the ground state, but does not yet specify their
order.

However, let us now view this ring of six colloids around
\texttt{A} as a periodic line. Figure \ref{fig:patterns-interactions}b (right) shows
that the $n(2)$-interactions on the triangular lattice also
correspond to the next-nearest neighbor interactions along this
ring. The implication is that the one-dimensional argument for
designability carries over directly to this situation, meaning
that the sequence \texttt{BCDEFG} will reliably form around
colloid \texttt{A}, up to a global rotation or reflection. By
iterative application of this argument to the points in the outer
ring of the hexagon (\texttt{B}, \texttt{C}, etc.), the proof
scales to arbitrarily large systems. Note that the orientation of
the subsequent hexagons is completely fixed with respect to the
initial patch. There are only \emph{global} degrees of freedom to orient the
pattern.

\subsubsection{Square lattice in two dimensions}

On the square lattice the self-repulsion energy $\varepsilon$ of
the design recipe should be extended to the $n(3)$ range (the
$J^{(3)}$ matrix). The proof then follows along the same lines as
that for the triangular lattice. We pick any lattice location and
consider the 8-vertex square around it (see figure \ref{fig:patterns-interactions}c). This square can be
represented as a line with periodic boundary conditions and
next-nearest neighbor interactions, proving the local uniqueness of the
ground state. By repeating this process iteratively for points on
the edge of the square the proof is extended to arbitrarily large
systems.

\section{Crystal structures}\label{sec:crystals}

We have shown that the design recipe produces a unique ground
state locally, and, by extension, for larger patches consisting
entirely of unique colloids. Of course, the ability to construct a
macroscopic materials that must consist of colloids with unique
DNA sequences is of limited use. Rather, one would like to use a
limited number of colloids to form a complex crystal that can grow
to arbitrary size. In this section, we show that the design recipe
can be used to construct various crystal structures.

Every crystal structure exhibits symmetries; at least two
independent translations. The 17 crystallographic \emph{wallpaper
groups} provide an exhaustive set of all such possible symmetry
groups on the two-dimensional plane \cite{Gruenbaum1987}. It
should be noted that a crystal structure can only exist on our
lattice model if its symmetry group is compatible with that of the
lattice itself. Specifically, the 5 wallpaper groups containing a
3-fold rotation are not compatible with a square lattice and the 3
groups with a 4-fold rotation are not compatible with the
triangular lattice. Formally, the symmetry group of the crystal
should be a subgroup of that of the lattice (\emph{p4m} or
\emph{p6m}), and at least contain two independent translations on
the lattice (\emph{p1}). For a pattern with symmetry group $g$, we
have $\mathit{p1} \subseteq g \subseteq \mathit{p4m}$ (square) or
$\mathit{p1} \subseteq g \subseteq \mathit{p6m}$ (triangular).

\begin{figure*}[htp]
\begin{center}
\includegraphics[width=0.9\textwidth]{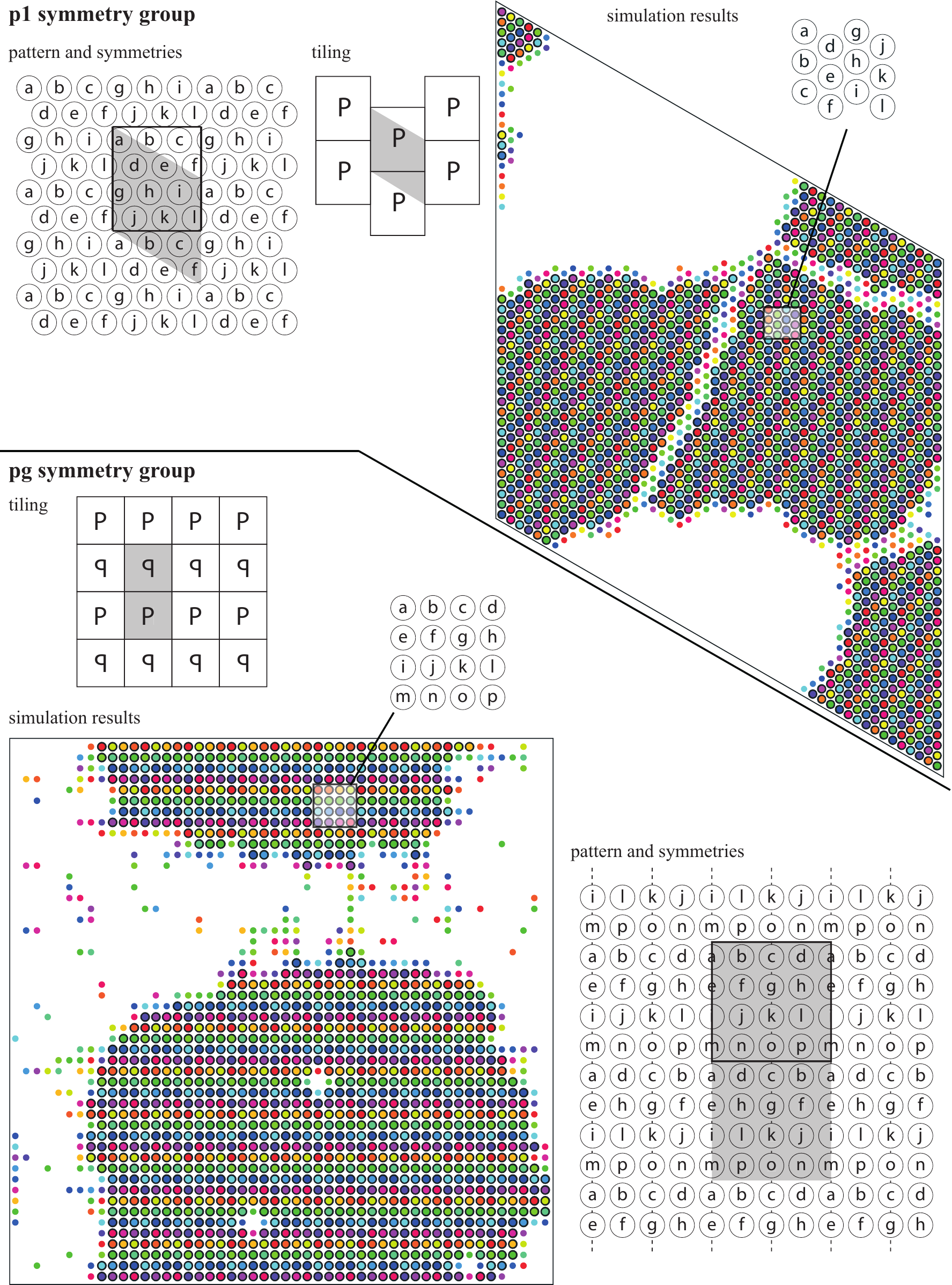}
\caption{Examples of a simple periodic crystal on a triangular
lattice (top; group \emph{p1}) and one with glide symmetries
(dashed lines) on a square lattice (bottom; group \emph{pg}). The
basic tile of colloids is delineated by the black rectangle, and
the unit cell is indicated by the gray parallelogram/rectangle.
The resulting tilings are shown with a P in each tile to indicate
its intrinsic orientation. In the simulation results, colloids
with locally ordered nearest-neighbors are indicated with a dark
ring.} \label{fig:result-p1-pg}
\end{center}
\end{figure*}

The interaction recipe outlined above can be successfully applied
to generate patterns representing 4 out of the 17 wallpaper groups
by selectively enabling symmetries. The simplest wallpaper group,
p1, represents a periodic crystal with only translation
symmetries. The ground state for such a pattern can be reliably
designed simply by instructing the colloids on one side of a patch
to connect to those on the opposite side. There is, however, a
lower limit to the translation distances that can be used. The
proof for the existence of a unique ground state in section
\ref{sec:groundstate} assumes that the colloids that form a ring
around any colloid \texttt{A} (see figure \ref{fig:patterns-interactions}b)
do not have attractive interactions with the colloids on the
opposite side of the ring. However, if one of the two translation
vectors becomes so short that the ring of colloids around colloid
\texttt{A} has nearest-neighbor interactions with another copy of
itself, the design recipe leads to spurious attractions that can
cause a degeneracy of the ground state. An example of a regular
periodic pattern on a triangular lattice is given in figure
\ref{fig:result-p1-pg} (top). Note that this pattern has the
shortest repeat length that is possible without creating spurious
interactions within a single colloid's local neighborhood.

\begin{figure*}[ht]
\begin{center}
\includegraphics{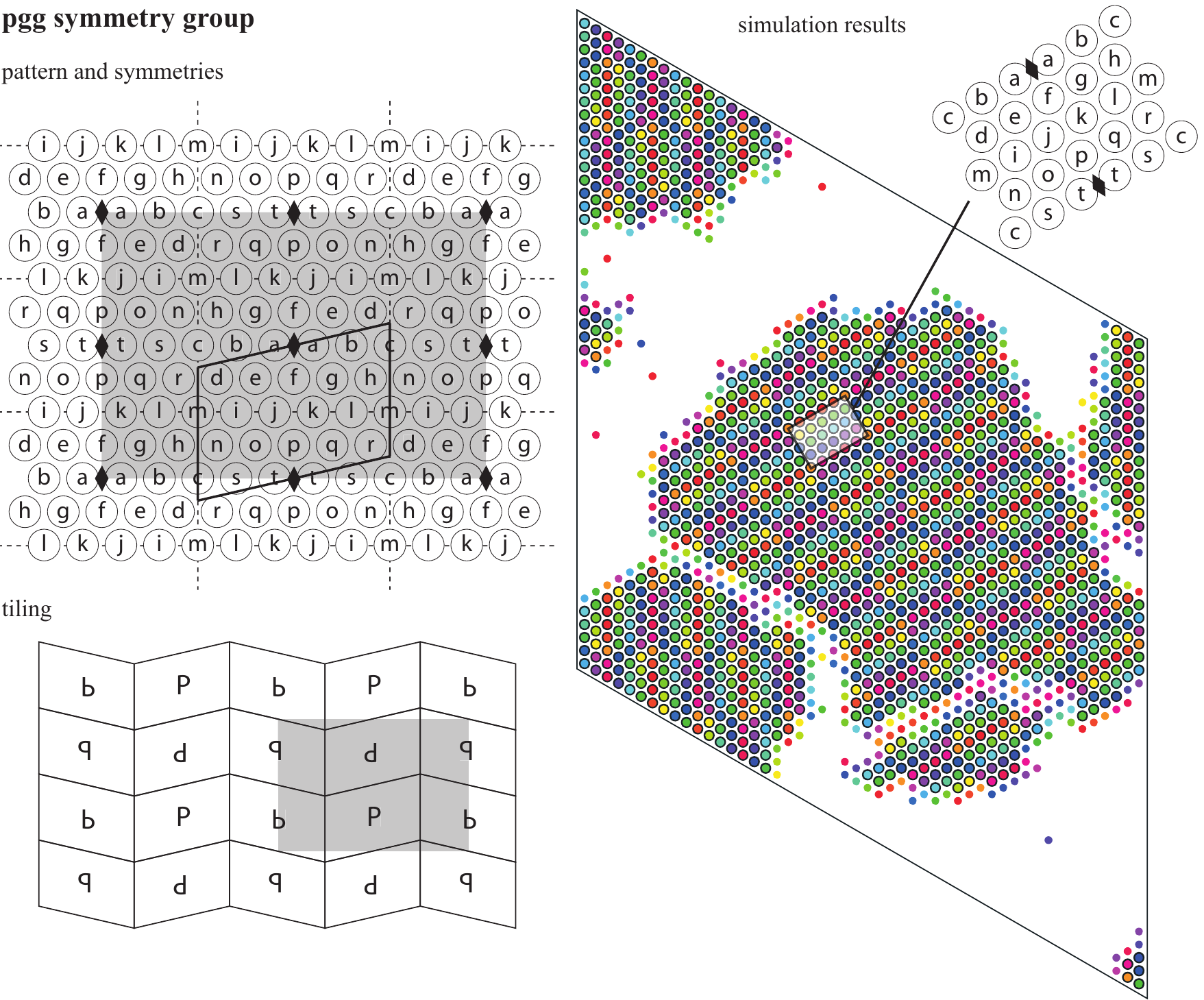}
\caption{Example of a periodic crystal (group \emph{pgg}) with
2-fold rotational (diamonds) and glide symmetries (dashed lines)
on a triangular lattice. The basic tile of colloids is delineated
by the black parallelogram, and the unit cell is indicated by the
gray rectangle. The resulting tiling is shown with a P in each
tile to indicate its intrinsic orientation. In the simulation
results, colloids with locally ordered nearest-neighbors are
indicated with a dark ring.} \label{fig:result-pgg}
\end{center}
\end{figure*}

Another symmetry that can be implemented in a straightforward
fashion is the glide reflection: a reflection followed by a
translation along the reflection axis. Again, if the translation
is large enough so that the local neighborhoods of identical
colloids do not touch one another, the glide reflection is simply
achieved by applying the interaction recipe to the desired
pattern. Producing a periodic pattern based on the glide
reflection generates the \emph{pg} wallpaper group, an example of
which is shown in figure \ref{fig:result-p1-pg} (bottom). For this
symmetry group, the unit cell is twice the size of the fundamental
domain, the patch of uniquely designed colloids.

As a final step we implement a non-trivial symmetry operation that
forces us to revisit the uniqueness proof of section
\ref{sec:groundstate}. This symmetry operation is the two-fold
rotation around the center point of two colloids on the triangular
lattice. An example of the local target pattern for this symmetry
is given in figure \ref{fig:patterns-interactions}d. For all the colloids
outside this region, the local neighborhood is qualitatively
unaffected, so the (local) ground state remains unique. For each
of the colloids \emph{inside} the region, one can enumerate the
possible local neighborhoods and verify that the target
configuration is the only one that corresponds to attractive
interactions on all lattice links and is therefore a good
candidate for the ground state. There is, however, an energy
penalty associated with the presence of two \texttt{B} colloids at
a next-nearest neighbor distance from each other. As long as this
energy penalty is similar in magnitude to the attractive
interactions -- and it's likely to be much smaller in practical
applications -- this penalty is outweighed by the attractive
interactions and the target pattern represents the unique ground
state. Obviously, a very large repulsive energy between the
B-colloids will disrupt the target pattern, but the exact
threshold value will generally depend on non-local interactions.
The use of a two-fold rotation allows us to generate patterns from
two other wallpaper groups: the group \emph{p2} with only two-fold
rotation symmetries (and resulting translations) and the group
\emph{pgg} that combines two-fold rotations with glide
reflections. An example of the latter is given in figure
\ref{fig:result-pgg}. For the group \emph{pgg}, the unit cell is
four times the size of the fundamental domain.

\section{Simulation results}

To illustrate our findings, we have run Monte Carlo
simulations starting from random initial conditions for the target
patterns shown in figures \ref{fig:result-p1-pg} and
\ref{fig:result-pgg}. Snapshots of the resulting configurations
are shown alongside the patterns. These clearly indicate a strong
tendency to assemble into the designed target patterns.

For the simulations, the interaction energies were chosen
according to the interaction recipe described above, with
$\alpha=3 k_B T$, $\varepsilon = 3 k_B T$. The system dimensions
were $50 \times 50$, forming a square for a square lattice and a
parallelogram for a triangular lattice, both with periodic boundary
conditions between opposing edges. The systems were initialized
with 2/3 of the sites occupied by colloids of all types in equal
proportions, and at random positions. For the evolution of the
system a Metropolis algorithm was used with $10^9$ steps in which
a swap of two random colloids was attempted. In all cases, this
led to sufficient (local) equilibration of the system. Local order was determined
by detecting whether a colloid had all the correct nearest
neighbors in the appropriate order. Colloids with locally ordered
neighborhoods are indicated by a black ring around them. Connected
regions of these colloids therefore indicate macroscopic crystal
structures of the correct type. In figures
\ref{fig:result-p1-pg} (top) and \ref{fig:result-pgg} distinct
grain boundaries are visible between macroscopic ordered patches with
incompatible orientations.

\section{Discussion}

In this work we have introduced a two-dimensional lattice model
for the interactions between DNA-coated colloids. The model allows
for a multitude of colloid types that interact according to
pair-specific interaction energies that are isotropic and of
finite range. For this system we have shown that it is
possible to choose the interaction energies in such a way that a
predetermined periodic pattern is the unique ground state.

Essential for the establishment of a non-trivial pattern as a
unique ground state is the ability of the interactions between
colloids to suppress the symmetries of the underlying lattice.
This has lead to the derivation in section \ref{sec:rangelimits}
of a lattice-dependent minimum interaction range. On a square
lattice, interactions need to extend to the next-next-nearest
neighbors, whereas on a triangular lattice, interactions up to and
including the next-nearest neighbors are sufficient.

Building on this result, we have introduced a simple recipe for choosing
the interactions between colloid species that is minimal in terms
of the interaction range between particles and guarantees the
local uniqueness of the ground state. The prescription consists of
an attractive nearest-neighbor potential for particles that should
become nearest neighbors in the final pattern and a self-repulsion
at larger distances (up to the required interaction range) between
particles of the same type. This recipe has been shown to work for
simple periodic patterns, patterns with glide reflections and
patterns with two-fold rotations on a triangular lattice.

The choice of interaction energies that has been made for the
simple recipe is presented as a proof of concept and is certainly
not unique. This is especially true for the choice for the
long-range self-repulsion that does not contribute to the energy
of the ground state, but merely prevents lattice symmetries from
disrupting the pattern. Another valid choice would be to add an
offset to all long-range interaction energies, whilst maintaining
a slightly more repulsive (or less attractive) interaction for
colloids of the same type.

Any such adjustments to the interaction recipe will need to be
made in the light of the kinetic aspects of self-assembly, which
have been ignored in the context of this work. For example, the
ground state of a system, however well designed, may be
kinetically inaccessible. On the other hand, judiciously chosen
interactions could speed up the formation of the ground state
pattern, for example through a `staged' ordering process
\cite{Lukatsky2006}.

\begin{figure}[ht]
\begin{center}
\includegraphics{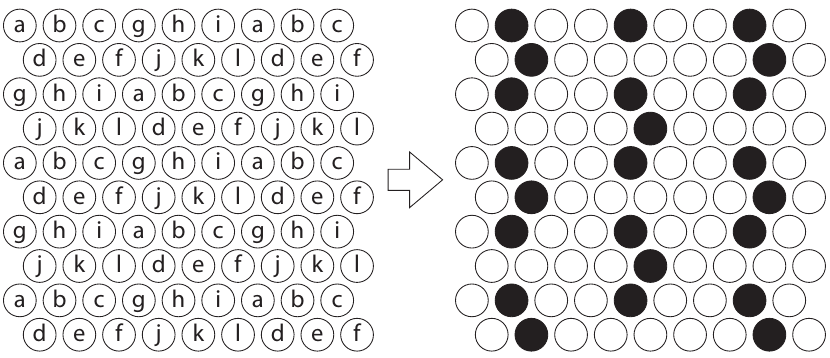}
\caption{Demonstration of the use of a periodic pattern of unique
colloids as a template. Colloids \texttt{b}, \texttt{e} and
\texttt{h} are colored black, the others are white.}
\label{fig:template}
\end{center}
\end{figure}

In section \ref{sec:crystals} various crystal structures have been
created by enabling specific symmetries: translations, glide
reflections and two-fold rotations. These crystals have
fundamental domains that consist of colloids that are all unique
in terms of their interactions. One of the distinct advantages of
more complex crystal structures over the simple periodic crystals
is that the fundamental domain, the basic unit of the tiling, is
smaller than the unit cell, limiting the number of uniquely
designed colloids that is required to make a pattern of a given
periodicity. For example, the unit cell of the \emph{pgg} symmetry
group is four times larger than its fundamental domain (see figure
\ref{fig:result-pgg}).

In addition, it is important to stress that the requirement for
all colloids in a tile to be unique only pertains to their
DNA-mediated interactions. The other properties of the colloids
can be chosen independently and can be used to create a higher
level pattern on top of the periodic crystal that now serves as a
template. For example, some colloids could have additional
chemical binding sites, or fluorescent or electrical properties.
The ability to use the underlying crystal as a template is
visualized in figure \ref{fig:template} for the simple periodic
pattern shown in figure \ref{fig:result-p1-pg} (top).

In future work, the interaction recipe presented in this work
could be extended to allow for additional symmetries, enabling
even more complex crystals to be designed. Furthermore, the
techniques that have been introduced in this work can be
applied to three-dimensional lattices, which have many more
symmetries, but are not fundamentally different from the
two-dimensional lattices discussed here.

\begin{acknowledgments}
We thank Koos van Meel for helpful suggestions on the manuscript.
This work is part of the research program of the ``Stichting voor
Fundamenteel Onderzoek der Materie (FOM)'', which is financially
supported by the ``Nederlandse organisatie voor Wetenschappelijk
Onderzoek (NWO)''.
\end{acknowledgments}


\begin{thebibliography}{15}
\expandafter\ifx\csname
natexlab\endcsname\relax\def\natexlab#1{#1}\fi
\expandafter\ifx\csname bibnamefont\endcsname\relax
  \def\bibnamefont#1{#1}\fi
\expandafter\ifx\csname bibfnamefont\endcsname\relax
  \def\bibfnamefont#1{#1}\fi
\expandafter\ifx\csname citenamefont\endcsname\relax
  \def\citenamefont#1{#1}\fi
\expandafter\ifx\csname url\endcsname\relax
  \def\url#1{\texttt{#1}}\fi
\expandafter\ifx\csname
urlprefix\endcsname\relax\def\urlprefix{URL }\fi
\providecommand{\bibinfo}[2]{#2}
\providecommand{\eprint}[2][]{\url{#2}}

\bibitem[{\citenamefont{Torquato}(2007)}]{Torquato2009}
\bibinfo{author}{\bibfnamefont{S.}~\bibnamefont{Torquato}},
  \bibinfo{journal}{Soft Matter} \textbf{\bibinfo{volume}{5}},
  \bibinfo{pages}{1157} (\bibinfo{year}{2007}).

\bibitem[{\citenamefont{Mirkin et~al.}(1996)\citenamefont{Mirkin, Letsinger,
  Mucic, and Storhoff}}]{Mirkin1996}
\bibinfo{author}{\bibfnamefont{C.~A.} \bibnamefont{Mirkin}},
  \bibinfo{author}{\bibfnamefont{R.~L.} \bibnamefont{Letsinger}},
  \bibinfo{author}{\bibfnamefont{R.~C.} \bibnamefont{Mucic}}, \bibnamefont{and}
  \bibinfo{author}{\bibfnamefont{J.~J.} \bibnamefont{Storhoff}},
  \bibinfo{journal}{Nature} \textbf{\bibinfo{volume}{382}}, \bibinfo{pages}{607
  } (\bibinfo{year}{1996}).

\bibitem[{\citenamefont{Leunissen et~al.}(2005)\citenamefont{Leunissen,
  Christova, Hynninen, Royall, Campbell, Imhof, Dijkstra, van Roij, and van
  Blaaderen}}]{Leunissen2005}
\bibinfo{author}{\bibfnamefont{M.~E.} \bibnamefont{Leunissen}},
  \bibinfo{author}{\bibfnamefont{C.~G.} \bibnamefont{Christova}},
  \bibinfo{author}{\bibfnamefont{A.-P.} \bibnamefont{Hynninen}},
  \bibinfo{author}{\bibfnamefont{C.~P.} \bibnamefont{Royall}},
  \bibinfo{author}{\bibfnamefont{A.~I.} \bibnamefont{Campbell}},
  \bibinfo{author}{\bibfnamefont{A.}~\bibnamefont{Imhof}},
  \bibinfo{author}{\bibfnamefont{M.}~\bibnamefont{Dijkstra}},
  \bibinfo{author}{\bibfnamefont{R.}~\bibnamefont{van Roij}}, \bibnamefont{and}
  \bibinfo{author}{\bibfnamefont{A.}~\bibnamefont{van Blaaderen}},
  \bibinfo{journal}{Nature} \textbf{\bibinfo{volume}{437}}, \bibinfo{pages}{235
  } (\bibinfo{year}{2005}).

\bibitem[{\citenamefont{Biancaniello et~al.}(2005)\citenamefont{Biancaniello,
  Kim, and Crocker}}]{Biancaniello2005}
\bibinfo{author}{\bibfnamefont{P.~L.} \bibnamefont{Biancaniello}},
  \bibinfo{author}{\bibfnamefont{A.~J.} \bibnamefont{Kim}}, \bibnamefont{and}
  \bibinfo{author}{\bibfnamefont{J.~C.} \bibnamefont{Crocker}},
  \bibinfo{journal}{Physical Review Letters} \textbf{\bibinfo{volume}{94}},
  \bibinfo{eid}{058302} (\bibinfo{year}{2005}).

\bibitem[{\citenamefont{Hill et~al.}(2008)\citenamefont{Hill, Macfarlane,
  Senesi, Lee, Park, and Mirkin}}]{Hill2008}
\bibinfo{author}{\bibfnamefont{H.~D.} \bibnamefont{Hill}},
  \bibinfo{author}{\bibfnamefont{R.~J.} \bibnamefont{Macfarlane}},
  \bibinfo{author}{\bibfnamefont{A.~J.} \bibnamefont{Senesi}},
  \bibinfo{author}{\bibfnamefont{B.}~\bibnamefont{Lee}},
  \bibinfo{author}{\bibfnamefont{S.~Y.} \bibnamefont{Park}}, \bibnamefont{and}
  \bibinfo{author}{\bibfnamefont{C.~A.} \bibnamefont{Mirkin}},
  \bibinfo{journal}{Nano Letters} \textbf{\bibinfo{volume}{8}},
  \bibinfo{pages}{2341} (\bibinfo{year}{2008}).

\bibitem[{\citenamefont{Nykypanchuk et~al.}(2008)\citenamefont{Nykypanchuk,
  Maye, van~der Lelie, and Gang}}]{Nykypanchuk2008}
\bibinfo{author}{\bibfnamefont{D.}~\bibnamefont{Nykypanchuk}},
  \bibinfo{author}{\bibfnamefont{M.~M.} \bibnamefont{Maye}},
  \bibinfo{author}{\bibfnamefont{D.}~\bibnamefont{van~der Lelie}},
  \bibnamefont{and} \bibinfo{author}{\bibfnamefont{O.}~\bibnamefont{Gang}},
  \bibinfo{journal}{Nature} \textbf{\bibinfo{volume}{451}}, \bibinfo{pages}{549
  } (\bibinfo{year}{2008}).

\bibitem[{\citenamefont{Park et~al.}(2008)\citenamefont{Park, Lytton-Jean, Lee,
  Weigand, Schatz, and Mirkin}}]{Park2008}
\bibinfo{author}{\bibfnamefont{S.~Y.} \bibnamefont{Park}},
  \bibinfo{author}{\bibfnamefont{A.~K.~R.} \bibnamefont{Lytton-Jean}},
  \bibinfo{author}{\bibfnamefont{B.}~\bibnamefont{Lee}},
  \bibinfo{author}{\bibfnamefont{S.}~\bibnamefont{Weigand}},
  \bibinfo{author}{\bibfnamefont{G.~C.} \bibnamefont{Schatz}},
  \bibnamefont{and} \bibinfo{author}{\bibfnamefont{C.~A.}
  \bibnamefont{Mirkin}}, \bibinfo{journal}{Nature}
  \textbf{\bibinfo{volume}{451}}, \bibinfo{pages}{553 } (\bibinfo{year}{2008}).

\bibitem[{\citenamefont{Geerts et~al.}(2008)\citenamefont{Geerts, Schmatko, and
  Eiser}}]{Geerts08}
\bibinfo{author}{\bibfnamefont{N.}~\bibnamefont{Geerts}},
  \bibinfo{author}{\bibfnamefont{T.}~\bibnamefont{Schmatko}}, \bibnamefont{and}
  \bibinfo{author}{\bibfnamefont{E.}~\bibnamefont{Eiser}},
  \bibinfo{journal}{Langmuir} \textbf{\bibinfo{volume}{24}},
  \bibinfo{pages}{5118} (\bibinfo{year}{2008}).

\bibitem[{\citenamefont{Tkachenko}(2002)}]{Tkachenko2002}
\bibinfo{author}{\bibfnamefont{A.~V.} \bibnamefont{Tkachenko}},
  \bibinfo{journal}{Physical Review Letters} \textbf{\bibinfo{volume}{89}},
  \bibinfo{pages}{148303} (\bibinfo{year}{2002}).

\bibitem[{\citenamefont{Lukatsky and Frenkel}(2004)}]{Lukatsky2004}
\bibinfo{author}{\bibfnamefont{D.~B.} \bibnamefont{Lukatsky}} \bibnamefont{and}
  \bibinfo{author}{\bibfnamefont{D.}~\bibnamefont{Frenkel}},
  \bibinfo{journal}{Physical Review Letters} \textbf{\bibinfo{volume}{92}},
  \bibinfo{eid}{068302} (\bibinfo{year}{2004}).

\bibitem[{\citenamefont{Bozorgui and Frenkel}(2008)}]{Bozorgui2008}
\bibinfo{author}{\bibfnamefont{B.}~\bibnamefont{Bozorgui}} \bibnamefont{and}
  \bibinfo{author}{\bibfnamefont{D.}~\bibnamefont{Frenkel}},
  \bibinfo{journal}{Physical Review Letters} \textbf{\bibinfo{volume}{101}},
  \bibinfo{eid}{045701} (\bibinfo{year}{2008}).

\bibitem[{\citenamefont{Licata and Tkachenko}(2006)}]{Licata2006}
\bibinfo{author}{\bibfnamefont{N.~A.} \bibnamefont{Licata}} \bibnamefont{and}
  \bibinfo{author}{\bibfnamefont{A.~V.} \bibnamefont{Tkachenko}},
  \bibinfo{journal}{Physical Review E} \textbf{\bibinfo{volume}{74}},
  \bibinfo{eid}{041406} (\bibinfo{year}{2006}).

\bibitem[{\citenamefont{Lukatsky et~al.}(2006)\citenamefont{Lukatsky, Mulder,
  and Frenkel}}]{Lukatsky2006}
\bibinfo{author}{\bibfnamefont{D.~B.} \bibnamefont{Lukatsky}},
  \bibinfo{author}{\bibfnamefont{B.~M.} \bibnamefont{Mulder}},
  \bibnamefont{and} \bibinfo{author}{\bibfnamefont{D.}~\bibnamefont{Frenkel}},
  \bibinfo{journal}{Journal of Physics: Condensed Matter}
  \textbf{\bibinfo{volume}{18}}, \bibinfo{pages}{S567} (\bibinfo{year}{2006}).

\bibitem[{\citenamefont{Crocker}(2008)}]{Crocker2008}
\bibinfo{author}{\bibfnamefont{J.~C.} \bibnamefont{Crocker}},
  \bibinfo{journal}{Nature} \textbf{\bibinfo{volume}{451}}, \bibinfo{pages}{528
  } (\bibinfo{year}{2008}).

\bibitem[{\citenamefont{Gr\"unbaum and Shephard}(1987)}]{Gruenbaum1987}
\bibinfo{author}{\bibfnamefont{B.}~\bibnamefont{Gr\"unbaum}} \bibnamefont{and}
  \bibinfo{author}{\bibfnamefont{G.}~\bibnamefont{Shephard}},
  \emph{\bibinfo{title}{Tilings and patterns}} (\bibinfo{publisher}{W.H.
  Freeman and company}, \bibinfo{year}{1987}), ISBN
  \bibinfo{isbn}{0-7167-1193-1}.

\end{thebibliography}
\end{document}